\documentclass[aps,prx,amsmath,amssymb,reprint,superscriptaddress]{revtex4-1}
\usepackage{graphicx}
\usepackage{amsmath}
\usepackage{dcolumn}
\usepackage{bm}
\usepackage{bbold}
\usepackage{color}
\usepackage{amsfonts}
\usepackage{amssymb}
\usepackage{mathrsfs}
\usepackage{tabularx}
\usepackage{braket}
\usepackage{mathtools}
\usepackage{soul}			

\usepackage{caption}
\usepackage{subcaption}

\begin{document} 
\title{High Cooperativity Cavity QED with Magnons at Microwave Frequencies}

\author{Maxim Goryachev}
\affiliation{ARC Centre of Excellence for Engineered Quantum Systems, School of Physics, University of Western Australia, 35 Stirling Highway, Crawley WA 6009, Australia}

\author{Warrick G. Farr}
\affiliation{ARC Centre of Excellence for Engineered Quantum Systems, School of Physics, University of Western Australia, 35 Stirling Highway, Crawley WA 6009, Australia}

\author{Daniel L. Creedon}
\affiliation{ARC Centre of Excellence for Engineered Quantum Systems, School of Physics, University of Western Australia, 35 Stirling Highway, Crawley WA 6009, Australia}

\author{Yaohui Fan}
\affiliation{ARC Centre of Excellence for Engineered Quantum Systems, School of Physics, University of Western Australia, 35 Stirling Highway, Crawley WA 6009, Australia}

\author{Mikhail Kostylev}
\affiliation{Magnetisation Dynamics and Spintronics Group, School of Physics, University of Western Australia, 35 Stirling Highway, Crawley WA 6009, Australia}

\author{Michael E. Tobar}
\email{michael.tobar@uwa.edu.au}
\affiliation{ARC Centre of Excellence for Engineered Quantum Systems, School of Physics, University of Western Australia, 35 Stirling Highway, Crawley WA 6009, Australia}

\date{\today}


\begin{abstract}

Using a sub-millimetre sized YIG (Yttrium Iron Garnet) sphere mounted in a magnetic field-focusing cavity, we demonstrate an ultra-high cooperativity of $10^5$ between magnon and photon modes at millikelvin temperatures and microwave frequencies. The cavity is designed to act as a magnetic dipole by using a novel multiple-post approach, effectively focusing the cavity magnetic field within the YIG crystal with a filling factor of 3\%. Coupling strength (normal-mode splitting) of 2~GHz, (equivalent to 76 cavity linewidths or $0.3$~Hz per spin), is achieved for a bright cavity mode that constitutes about 10\% of the photon energy and shows that ultra-strong coupling is possible in spin systems at microwave frequencies. With straight forward optimisations we demonstrate that with that this system has the potential to reach cooperativities of $10^7$, corresponding to a normal mode splitting of 5.2 GHz and a coupling per spin approaching 1 Hz. We also observe a three-mode strong coupling regime between a dark cavity mode and a magnon mode doublet pair, where the photon-magnon and magnon-magnon couplings (normal-mode splittings) are 143~MHz and 12.5~MHz respectively, with HWHM bandwidth of about 0.5~MHz.
\end{abstract}

\maketitle

\section*{Introduction}

The field of quantum information has achieved significant progress in recent decades, and its continued success is driven by a strong focus on quantum technology. Ongoing development of the devices and experimental techniques that utilize the quantum nature of the world to perform storage, transfer and processing of quantum information is critical to achieving the ambitious goal of workable quantum computation. The mainstream framework used to reach this goal is known as Quantum Electrodynamics (QED), having roots in the Jaynes-Cummings model and its variations\cite{JC,Tavis:1968kq}. Under this framework, a high-fidelity technology must be able to exchange information with preserved coherence, i.e. to demonstrate the so-called strong coupling regime, which is represented by a cooperitivity greater than unity. This criteria, which must be met in order to be useful for any quantum application, is characterised by a coupling between two subsystems that is stronger than the mean of the losses in both of them. In the present work, this is the regime in which the photon-magnon coupling is greater than the average of the resonant cavity and resonant magnon losses.

Achieving operation in a strong coupling regime is a challenging task, because one usually encounters contradictory requirements for the coupling of a system to its environment. In this work, we push the limits of what is currently possible and achieve extremely high cooperitivity at microwave frequencies. Such cooperitivities have only been achieved previously in optical systems coupled to the motion of neutral atoms\cite{PhysRevLett.99.213601}. The unique nature of the system described herein allows the `ultra-strong' coupling regime to be approached, in which the coupling energy is comparable to that of the subsystem itself\cite{Ciuti:2006vn,Niemczyk:2010kx,Plumridge:2008ys,Devoret:2007zr}. This relation between the energy of the coupled and uncoupled system invalidates the use of the usual Rotating Wave Approximation, and thus leads to a breakdown of the Jaynes-Cummings model. The resulting system is analytically unsolvable and demonstrates complex dynamics that may include frequency renormalization, revival and collapse of coherent oscillations, and chaotic behavior\cite{PhysRevA.85.043815,PhysRevB.72.195410}. Until now the ultra-strong coupling regime has only been observed in certain systems at optical frequencies \cite{Schlather:2013ff,Kena-Cohen:2013aa,Schwartz:2011aa,Todorov:2010aa,Scalari}, in artificially created matter, and with superconducting qubits\cite{Niemczyk:2010kx,Devoret:2007zr}. The present work demonstrates that with a specially designed photonic cavity, this regime may also be achieved in a sub-millimetre sized spin system at microwave frequencies. Previous work\cite{Schlather:2013ff} has achieved ultra-strong `light-matter' coupling through electric near-field enhancement using an electric dipole. The present work is the magnetic analogue of this, instead using a magnetic dipole configuration to enhance the field within the sample. 

The first step towards a new physical realisation of a scheme for manipulating quantum information requires that a proper choice of subsystems be made. Whilst one of the subsystems is typically an electromagnetic cavity, the other must demonstrate quantum behaviour, preferably with long coherence times. Examples of the latter subsystem are numerous and range from mechanical oscillators in their quantum ground state to trapped single ions. For the former subsystem, a choice is typically made between a 2D or 3D electromagnetic cavity. Despite recent progress in 2D planar superconducting structures, the use of 3D structures has gained broad interest in recent years and continues to gain momentum\cite{Paik:2011aa,Rigetti:2012aa}. 

A number of physical realisations have been proposed recently for the `quantum' subsystem that include trapped ions\cite{Gessner:2014kn}, superconducting qubits\cite{Wallraff:2004tw}, quantum dots\cite{Reithmaier:2004qo}, photonic nanostructures\cite{Sato:2012ye}, ultra-cold atoms and ensembles\cite{ritsch}, and electronic and nuclear spins in solids\cite{PhysRevLett.107.060502}. In particular, one possible realisation of an optical-to-microwave quantum interface \cite{PhysRevLett.109.130503} is based on spin-doped dielectric crystals such as Fe$^{3+}$ ions in sapphire\cite{PhysRevB.88.224426}, Er$^{3+}$ in yttrium orthosilicate (YSO)\cite{Bushev:2011ve, Probst:2013zg}, or NV-centres in diamond\cite{Gaebel:2006fu}.  The use of large spin ensembles for quantum information processing has been discussed previously by Imamoglu \cite{PhysRevLett.102.083602} who claims that collective excitations of spin ensembles can be incorporated in hybrid systems with a nonlinear element such as a Josephson junction. This includes the possibility to design new protocols without single-spin confinement \cite{PhysRevLett.102.083602}. In general, large spin ensembles are considered to be of great potential importance for application to hybrid quantum systems and quantum information manipulation, with considerable work invested in this area by many groups \cite{PhysRevLett.105.140503,PhysRevLett.107.060502,PhysRevLett.105.140502,Zhu:2011aa}.

 The utilisation of spin-doped 3D cavities can lead to operation in the strong coupling regime, however the coupling strength always stays limited because an increase in the number of spins usually leads to broadening of both the spin and cavity resonances causing a trade-off with electromagnetic coupling. This problem may be solved by using dielectrics that exhibit ferromagnetism instead of dilute paramagnetic impurities. Because ferromagnetic materials are perfectly ordered systems, they do not suffer from excess losses due to spin-spin interactions when the spin density is increased. At the same time, ferrite materials have much larger magnetic susceptibilities than paramagnetic systems due to the vastly increased number of spins, thus having the potential for much stronger cavity-spin coupling per unit volume. Furthermore, ferromagnetic magnon resonances could be compared to mechanical modes, another important element of optomechanics-based quantum hybrid systems\cite{Kippen}. A natural choice for a ferromagnetic system is single-crystal YIG, or yttrium iron garnet (Y$_3$Fe$_2$(FeO$_4$)$_3$)\cite{lvov}, a ferrite material with unique microwave properties. YIG exhibits a record low microwave magnetic loss parameter, and excellent dielectric properties at microwave frequencies. For this reason, it has been extensively studied at room temperature for various microwave and optical applications. Strong coupling regimes in YIG nanomagnets have already been predicted \cite{PhysRevLett.104.077202,Soykal:2010ly,PhysRevB.82.104413}, and although some preliminary attempts have been made to couple superconducting planar cavities\cite{Huebl:2013pi} and 3D resonators\cite{PhysRevLett.113.083603,yig2} to magnon resonances in YIG, the potential of the material has not yet been fully explored. Thus, the combination of a specially designed 3D cavity and a YIG ferromagnet represents a promising path towards the ultra-strong coupling regime of QED.

Specifically, in the present work we use a sub-millimetre sized YIG sphere mounted in a 3D microwave cavity at millikelvin temperatures and approach the ultra-strong coupling regime between magnon and photon modes in the system. The cavity is designed using a novel, patented\cite{patent2014}, multiple post re-entrant cavity concept. This configuration effectively focuses the resonant magnetic field into the sub-millimetre sized YIG crystal to achieve extraordinary large filling factors at microwave frequencies. Such a large magnetic filling factor is possible despite the fact that the smallest resonant frequency of the YIG crystal itself is of the order of 100 GHz. Coupling strength of 2~GHz is achieved for the bright cavity mode, which constitutes about $10\%$ of the photon energy, or nearly 76 cavity linewidths. In addition, a three-mode strong coupling regime is observed between a dark cavity mode and two magnon modes, where the photon-magnon and magnon-magnon couplings are 143~MHz and 12.5~MHz respectively. Multiple magnon modes of the YIG sphere are observed with bandwidths approaching 0.5~MHz.

\section*{Cavity with Magnetic Field Focusing}
\label{fillfact}

Due to the peculiar structure of magnon modes in a ferromagnetic material, the optimal shape of a 3D ferromagnetic resonator is a miniature sphere. As such, best commercial quality YIG crystals typically come in the form of spheres for microwave applications, however technological limitations place bounds on the maximum volume of such crystals. Single-crystal spherical YIG resonators can typically be manufactured with a diameter in the range 200--1000 $\mu$m. These dimensions make the corresponding resonant frequencies of electromagnetic modes in the sphere above 100 GHz. Thus, it is impossible to utilise the sphere itself as a photon cavity in the X and K$_\text{u}$ microwave frequency bands, and one must instead rely on coupling of the ferromagnetic resonance in the sphere to some external resonator, for example a 3D microwave cavity. For a traditional rectangular microwave cavity, the half wavelength of the lowest order standing wave resonance is equal to the cavity size. In the X and K$_\text{u}$ bands, the half wavelength is much greater than the diameter of the sphere, strongly reducing the coupling of the sphere to the cavity.  In the present work we use a novel design of re-entrant cavity with two posts to increase the filling factor of the cavity volume with the YIG material, and thus enhance the coupling. In comparison, recent work which couples standard cavities to similar YIG spheres obtained significantly lower filling factors, cooperitivity and coupling per spin\cite{PhysRevLett.113.083603,yig2}

A re-entrant cavity\cite{reen0,reen1,reen2} is a closed (typically cylindrical) 3D microwave resonator built around a central post with an extremely small gap between one of the cavity walls and the post tip. Such a cavity is characterised by spatial separation of the electric and magnetic components of the cavity field. Whereas most of the electric field is concentrated in the gap, all of the magnetic field is distributed around the post, with a fast radial decay in amplitude. In fact, a re-entrant cavity can be considered the 3D realisation of a lumped $LC$ circuit and thus can operate in a sub-wavelength regime.

\begin{figure*}[t!]
			\includegraphics[width=1\textwidth]{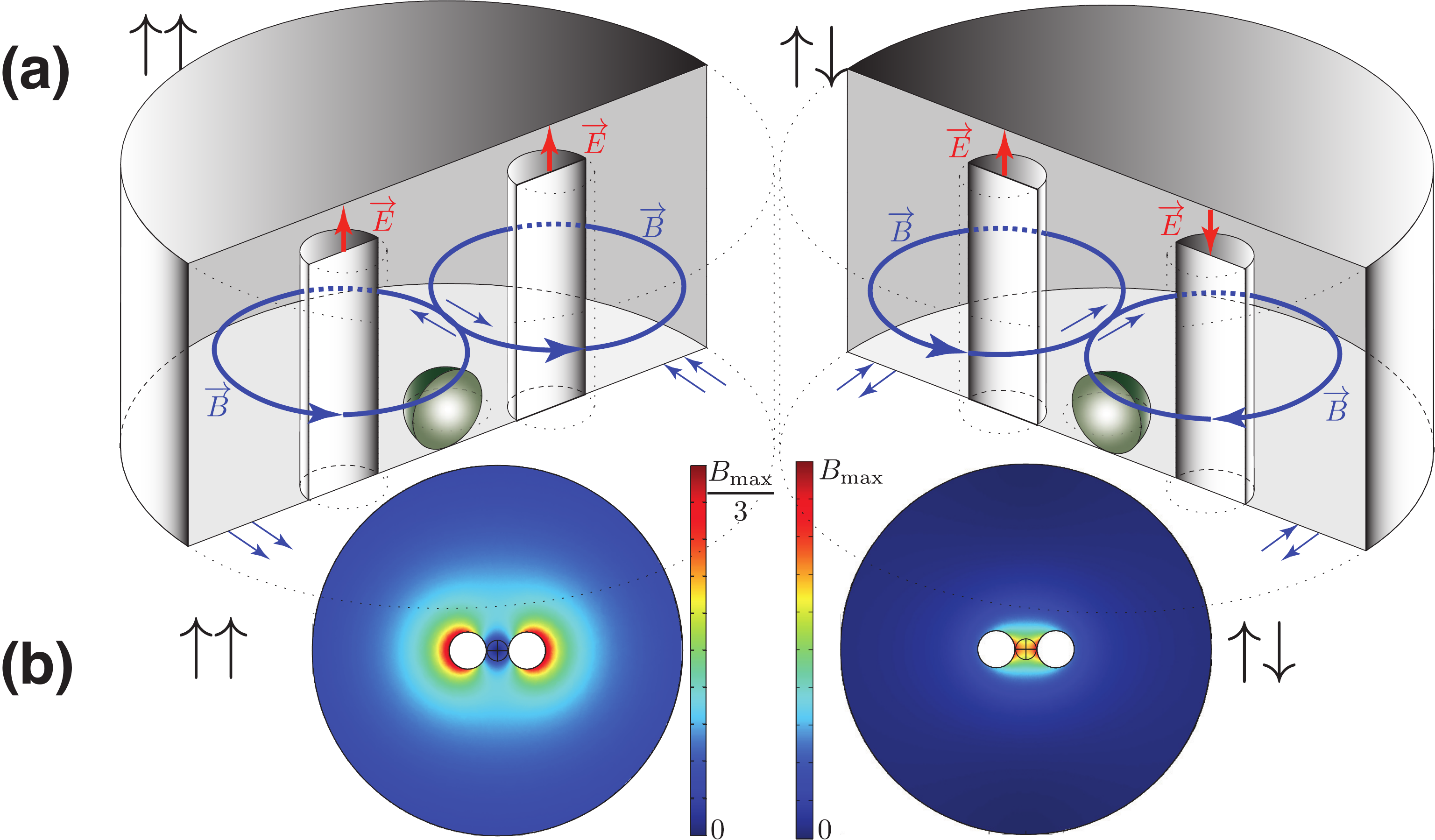}
	\caption{Two re-entrant type cavities with two resonance posts. Cavity $\uparrow\uparrow$  demonstrates operation in the `dark' mode, and cavity $\uparrow\downarrow$ demonstrates operation in the `bright' mode. (a) shows a 3D cross-sectional view of the cavities with $\vec{E}$ and $\vec{B}$ field directions, and (b) shows a top view of the field modelling results computed in the equatorial plane of the YIG sphere.}
	\label{cavity}
\end{figure*}

Microwave re-entrant cavities as described have a well confined electrical field, but the magnetic field is distributed over quite some volume which is significantly larger than that of a typical YIG sphere. Thus, it is necessity to focus this field into a relatively small region of the cavity, a problem that is solved by employing a double-post structure (see Fig.~\ref{cavity}, (a)) which is the simplest case of a 3D reentrant cavity lattice\cite{mpost}. Such multiple post re-entrant cavities are the subject of a patent\cite{patent2014}. In fact, each post acts as a separate microwave resonator since each has its own equivalent inductance and capacitance. As a result, the overall system has two modes: a dark mode, and a bright mode. An electrical analogue of this approach has been used previously to enhance electrical coupling between surface plasmons and molecular excitons\cite{Schlather:2013ff}.

For the dark mode (Fig.~\ref{cavity}, labelled $\uparrow\uparrow$) the current through both posts flows in the same direction, making the $\vec{B}$ field around each post curl in the same direction. As a result, the vectors of the $\vec{B}$ field from the two posts are anti-parallel in the space between the posts, effectively cancelling the field. Elsewhere, the field vectors are almost co-aligned, thus enhancing the field. The result of Finite Element Modelling of the magnetic field distribution for such a mode is shown in Fig.~\ref{cavity}(a). The model takes into account the presence of a YIG sphere at zero external applied magnetic field. Since all of the electric field is concentrated in the post gaps and no magnetic response is seen at zero-field, the presence of the sphere is a small perturbation of the solution. In the case of the bright mode, (Fig.~\ref{cavity}, labelled $\uparrow\downarrow$), the currents through the posts are always opposite, making the $\vec{B}$ field for each post curl in the opposite direction. As a result, the total $\vec{B}$ field is enhanced between the posts and cancelled elsewhere. The distribution of magnetic field for the bright mode is shown in Fig.~\ref{cavity}(b). The system described is similar to the magnetic field of a current loop. In the plane of the loop, the fields generated by any two opposite loop sections are added in phase within the bounded area of the loop, and out of phase outside the loop area. This strongly enhances the total field inside the loop and effectively cancels it outside. Thus, the cavity presented here can be understood as a two-dimensional current dipole. Note that for both cavity modes the microwave magnetic field lies perfectly in the plane perpendicular to the posts. 

The results of the field modelling shown in Fig.~\ref{cavity}(b) confirm that the presence of the two posts makes it possible to produce excellent confinement of the magnetic component of the cavity field in the small volume between the posts. This field-focusing effect results in very high spatial overlap between the photon mode of the cavity and the magnon mode of the YIG crystal, and thus the strong coupling between them. The overlap is usually characterised by a filling factor $\xi$ that denotes the portion of the total cavity magnetic energy stored in the sphere. Another important experimental parameter is the geometric factor, $G$, calculated as follows:
\begin{equation}
	\label{B006SFa}
	\displaystyle  G=\omega_{0}\mu_0\frac{\int_V |H|^2dv }{\int_S |H|^2ds},\\
\end{equation}
where $\omega_{0}$ is the angular frequency of a cavity mode, $\mu_0$ is the vacuum permeability, $H$ is magnetic field in the cavity and $V$ and $S$ are cavity volume and conducting surface.
This parameter relates the electromagnetic cavity $Q$-factor to the surface resistance $R_s$ through the expression $G=Q \times R_s$, assuming that this is the dominant loss mechanism.

\begin{figure*}[t!]
			\includegraphics[width=0.99\textwidth]{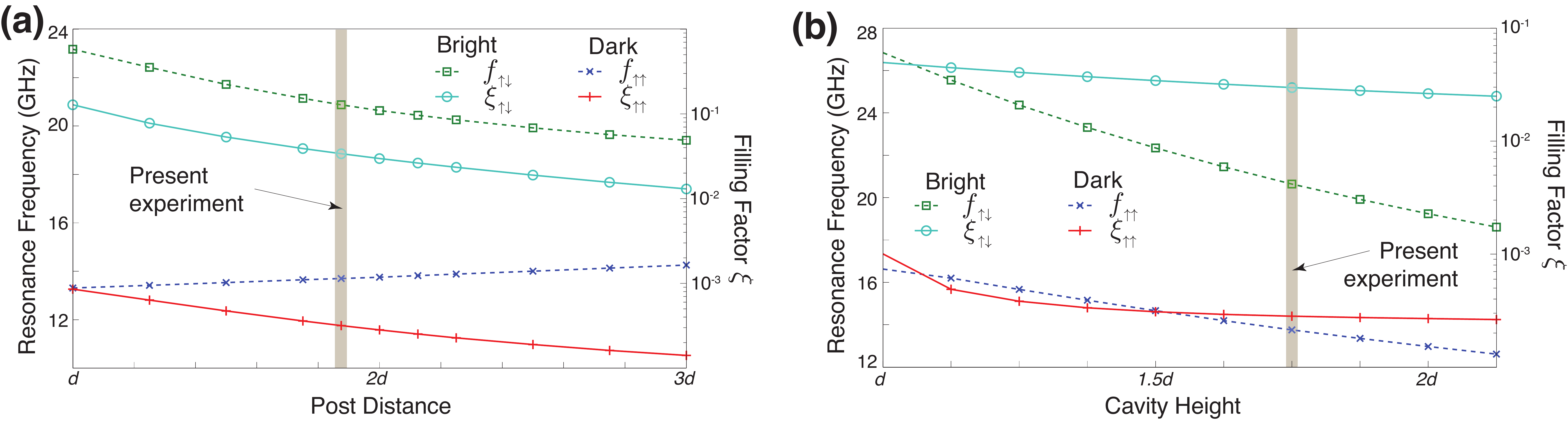}
	\caption{Cavity resonance frequencies and YIG sphere filling factors as a function of (a) the distance between the two posts and (b) the cavity height, for both the dark and bright modes. The distance is given in terms of the sphere diameter $d$.}
	\label{enhans}
\end{figure*}

Finite Element Modelling of the cavity modes allows estimations of the cavity eigenfrequencies, the filling factor $\xi$ and the geometric factor $G$. For the actual cavity dimensions (an internal cavity radius of 5~mm, cavity height $1.4$~mm, post radius of 0.4~mm, and post gap of 73~$\mu$m, distance between the posts $1.5$~mm) the predicted resonance frequencies for the dark and bright modes are $13.75$ and $20.6$GHz respectively, which is in good agreement with the experiment. The filling factors for the two modes are $3\times 10^{-4}$ and $3\times 10^{-2}$ respectively, a ratio of two orders of magnitude. The bright mode is also superior in terms of the geometric factor, 59 $\Omega$ versus 51 $\Omega$ for the dark mode. Note that the sphere is a small perturbation for the cavity mode when the magnon resonance is tuned away mainly due to the absence of the electric field between the two posts.

\begin{figure}[t!]
	\centering
			\includegraphics[width=0.5\textwidth]{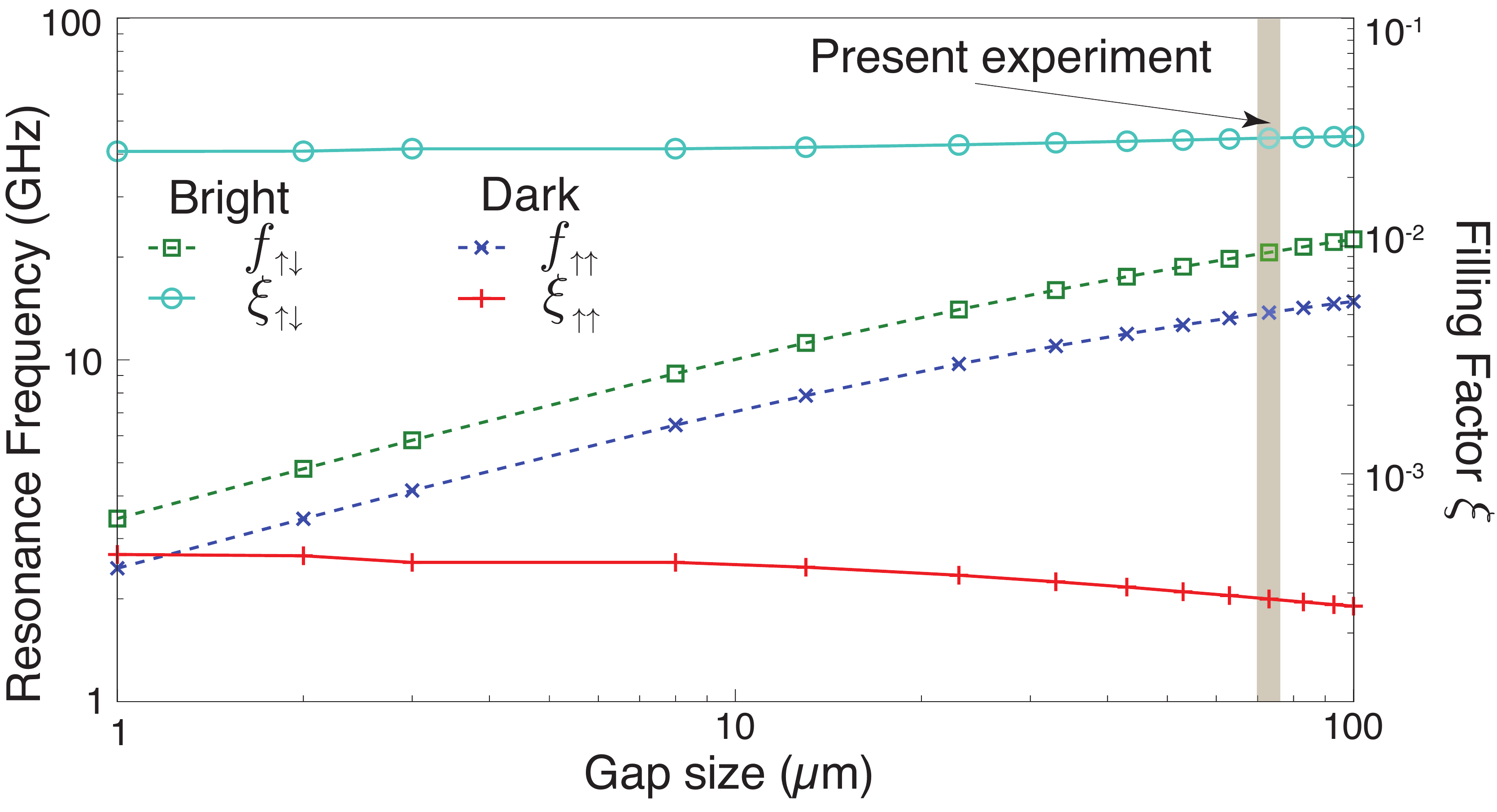}
	\caption{ Cavity resonance frequencies and YIG sphere filling factor as a function the post gap size, for both the dark and bright modes.}
	\label{gap}
\end{figure}

Our simulations demonstrate that the filling factor $\xi$ can be further enhanced by optimising the distance between the posts, and the height of the cavity. In Fig.~\ref{enhans}, the cavity resonance frequencies and filling factors are shown as a function of the corresponding dimensions in units of the sphere diameter $d=0.8$~mm. A decrease in post spacing to the diameter of the YIG sphere results in a filling factor increase up to $0.12$, whereas reducing the cavity height to the smallest possible value results in a filling factor of $0.05$. Applying both optimisations together would result in a filling factor of $0.2$, potentially an order of magnitude larger than the present work. Altering these parameters, however, results in an increase of the cavity eigenfrequencies. This drawback can be overcome by adjusting the size of the post gap, to which the cavity resonance frequency is extremely sensitive (Fig.~\ref{gap}) but the filling factor $\xi$ is completely insensitive. Reducing the gap size leads not only to a decrease in the cavity eigenfrequency, but also to an increase of the geometric factor. Two other major parameters, namely the post radius and the cavity diameter do not significantly change the value of the filling factor, although they can be also used to manipulate the resonant frequency.

\section*{Physical Realisation}

The current dipole cavity is fabricated from Oxygen-Free High Conductivity (OFHC) copper with the dimensions corresponding to that used for simulation in the previous section. After insertion of the YIG sphere between the posts, the cavity is cooled to about $25$ mK by means of a Dilution Refrigerator (DR) with a cooling power of about $500~\mu$W at 100 mK. The cavity is attached to a OFHC copper rod bolted to the mixing chamber stage of the DR that places it at the field centre of a $7$ T superconducting magnet. The magnet is attached to the 4 K stage of the DR, with the cavity mounted within a $\sim100$~mK radiation shield that sits within the bore of the magnet. 

A commercially available YIG sphere with a diameter $d=0.8$~mm attached to a standard beryllium oxide cylindrical post was used for the experiment. The nominal magnetic losses specified by the manufacturer are 0.2~Oe at room temperature. A cylindrical hole was made centrally in the bottom of the cavity in order to hold the sphere by its beryllium oxide mounting rod in between the posts. The optimum crystallographic axis for thermal compensation of the sphere is oriented along the posts. Since no thermal stability characterisation of the magnon resonance is made in this study, the orientation of the sphere is not important for the present work.

\begin{figure}[t!]
	\centering
			\includegraphics[width=0.5\textwidth]{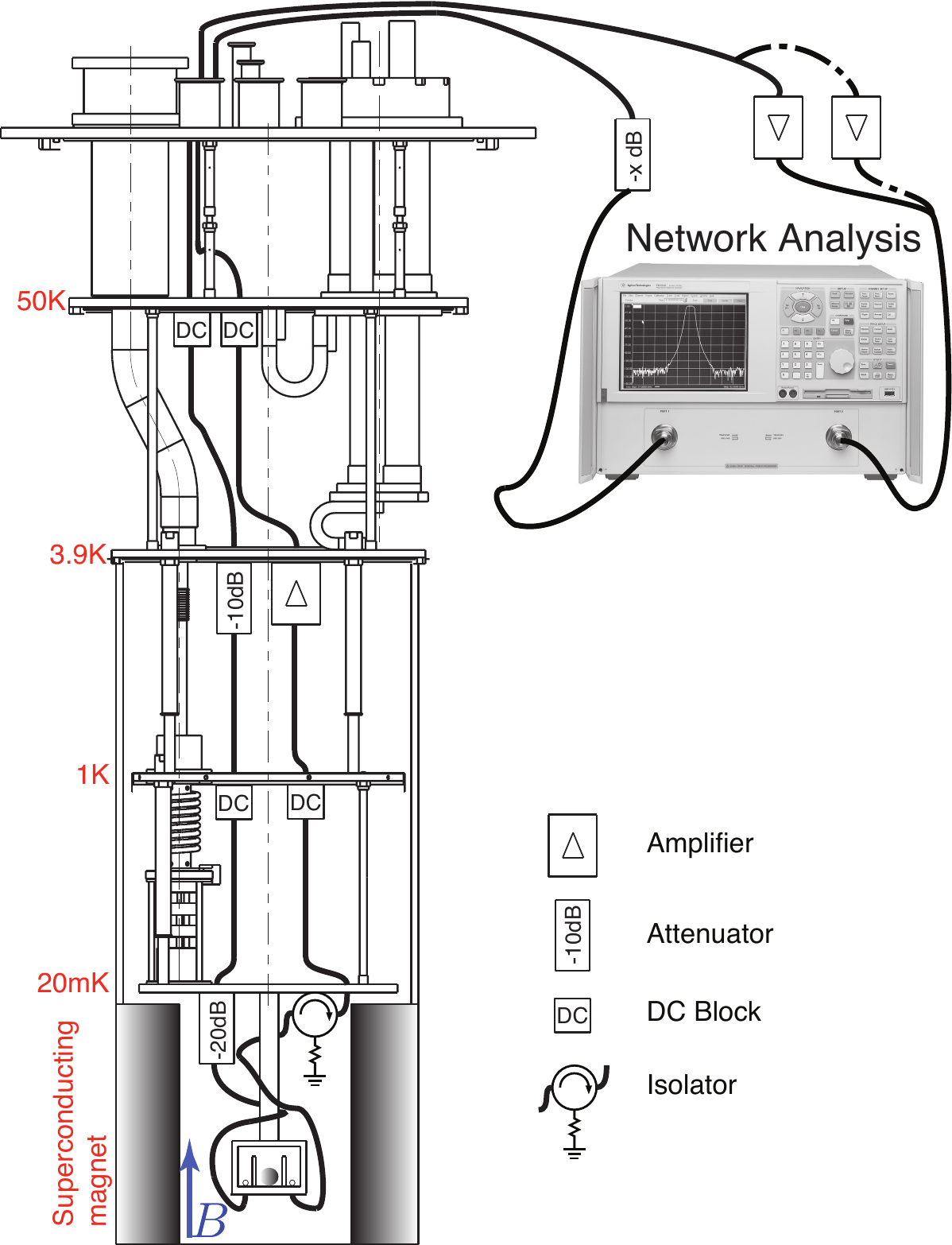}
	\caption{Schematic of the experiment. For clarity, not all anti-radiation shields are shown.}
	\label{exper}
\end{figure}

The cavity modes are exited by a loop probe constructed from flexible SMA cable launchers, and measurements are performed through a second loop probe. The incident signals are attenuated by a series of cold attenuators at 3.9 K ($-10$~dB) and at 20 mK ($-20$~dB) before reaching the cavity. The power incident on the cavity is set at $-90$ dBm, while the coupling on both ports does not exceed $0.01$ (corresponding to less than 15 photons in the cavity). The transmitted signal is then amplified by a cryogenic low noise amplifier (Low Noise Factory $6-20$~GHz, $20$ dB gain) bolted to the 3.9~K stage. The cryogenic amplifier and the cavity are separated by an isolator situated at the $20$~mK stage to prevent back action noise from the higher temperature stage. The isolator is shielded from the field of the superconducting magnet. Additional room temperature amplification is performed via one of two low noise amplifiers ($6-18$~GHz and $18-40$~GHz), dependent on the frequency band under study. Additional DC blocks are used at various stages of the system in order to suppress spurious low frequency signals. Both the network analyser and the superconducting magnet are computer controlled for automatic operation. Figure~\ref{exper} shows a schematic of the experimental setup.

Note that the superconducting magnet and the cavity are separated by an additional anti-radiation shield attached to the 100~mK stage (see Fig. ~\ref{exper}). The cavity is firmly connected to the 20~mK stage via a thick Oxygen Free High Conductivity copper rod. The cavity is placed at the centre of the superconducting magnet where sufficient uniformity of the DC magnetic field is achieved. 

\section*{Experimental Observations}
\label{observ}

The cavity was placed under an applied external DC magnetic field swept from 0 to 0.9 T, and its transmission response was recorded using a vector network analyser. The maximum driving efficiency of magnon modes by a microwave field is achieved when the field is perpendicular to the static magnetic bias field\cite{gurevich}. Accordingly, since the microwave magnetic field vector is in the plane normal to the posts, the DC magnetic field was oriented parallel to the cavity posts. 

\begin{figure*}[t!]
			\includegraphics[width=1\textwidth]{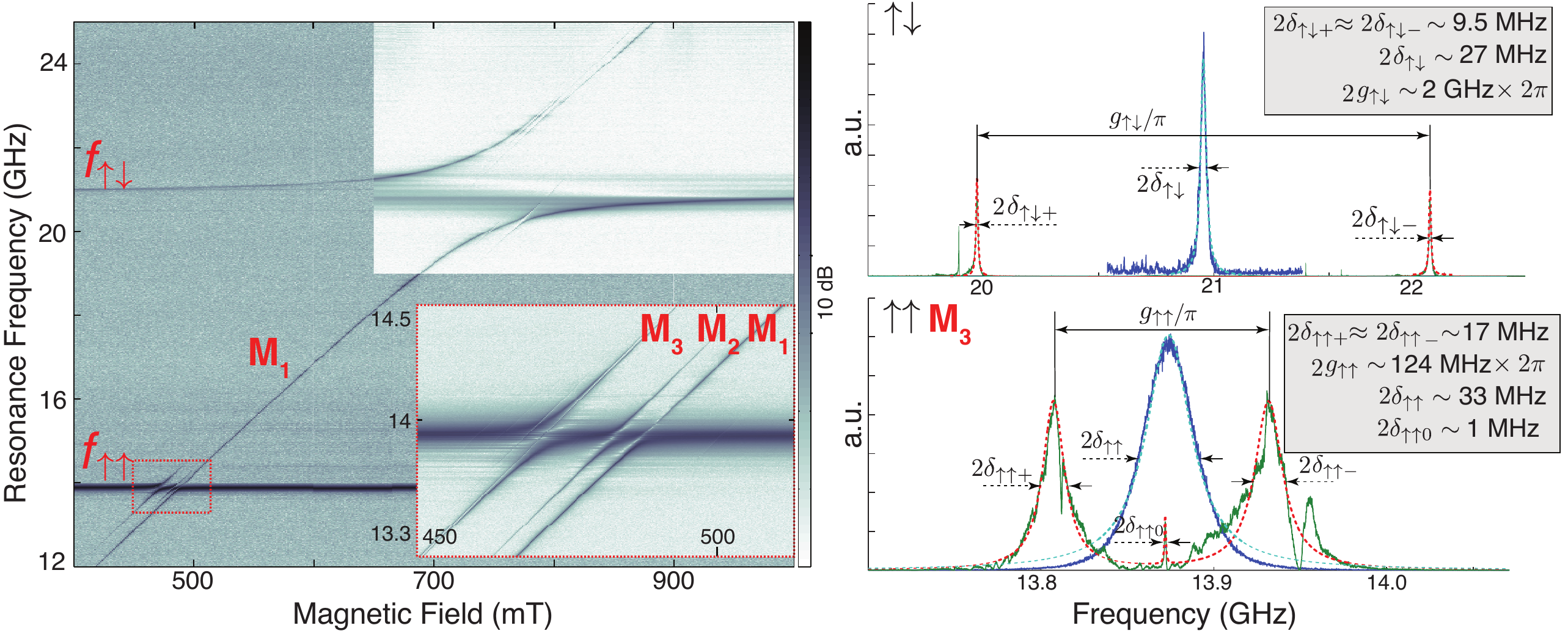}
	\caption{Transmission through the cavity as a function of frequency and applied DC magnetic field. The inset demonstrates the region of interaction between the dark mode and the magnon resonance. The plot labelled $\uparrow\downarrow$ shows the frequency response of the interaction between the $f_{\uparrow\downarrow}$ cavity mode and the magnon mode M$_1$ at $B=$ 0.743~T, and the bare cavity mode outside resonance. The plot labelled $\uparrow\uparrow$ shows the interaction between $f_{\uparrow\uparrow}$ and the magnon mode M$_2$ at $B=$ 0.471~T, as well as the bare cavity mode outside the resonance. Dashed curves represent Lorentzian fits to the data.}
	\label{density}
\end{figure*}

Figure~\ref{density} demonstrates the microwave response of the cavity with a YIG sphere located between the posts. The response is shown as a function of external magnetic field applied parallel to the posts, with darker colour corresponding to higher transmission. The upper right section of the density plot has a different signal-to-noise ratio than the rest of the data due to the utilisation of a different (higher frequency bandwidth) room temperature amplifier in that region. The two horizontal lines of higher transmission labelled as $f_{\uparrow\uparrow}=$13.9~GHz and $f_{\uparrow\downarrow}=$20.9~GHz correspond to the dark and bright cavity modes respectively. The line of higher transmission that grows almost linearly with the magnetic field represents the magnon mode of uniform precession in the YIG sphere. In addition to the mode labelled M$_1$, there are a number of other modes, so-called magnetostatic modes of the ferromagnetic sphere\cite{fletcher,Wiese} visible only in the close vicinity of the cavity resonances. The most prominent of these are labelled M$_2$ and M$_3$, and are clearly seen in the inset of the density plot in Fig.~\ref{density}. A darker region inside the upper interaction is due to line resonances that are removed elsewhere with image post-processing.

The density plot reveals the existence of several avoided crossings between the magnon modes of the sphere and the two cavity modes. The strongest interaction is observed between the bright cavity resonance and the uniform magnon mode. The corresponding strong coupling regime achieved at $B=$ 0.743~T is demonstrated in the sub-figure labelled $\uparrow\downarrow$. This plot displays a central peak representing the bare cavity resonance outside the interaction, with linewidth $2\delta_{\uparrow\downarrow}$ (given in units of Hz), and the splitting between the two peaks in the strong coupling regime (with linewidths $2\delta_{\uparrow\downarrow+}$ and $2\delta_{\uparrow\downarrow-}$). {The condition on the strong coupling is $\frac{g_{\uparrow\downarrow}}{\pi}\gg \delta_{\uparrow\downarrow+}+\delta_{\uparrow\downarrow-}$ is thus satisfied.} The strength of the coupling $g_{\uparrow\downarrow}/\pi$ is approximately 2~GHz, which is $10$\% of the corresponding resonance frequency $f_{\uparrow\downarrow}$, qualifying it as ultra-strong coupling as discussed in the introduction. 

\begin{figure*}[t!]
			\includegraphics[width=0.96\textwidth]{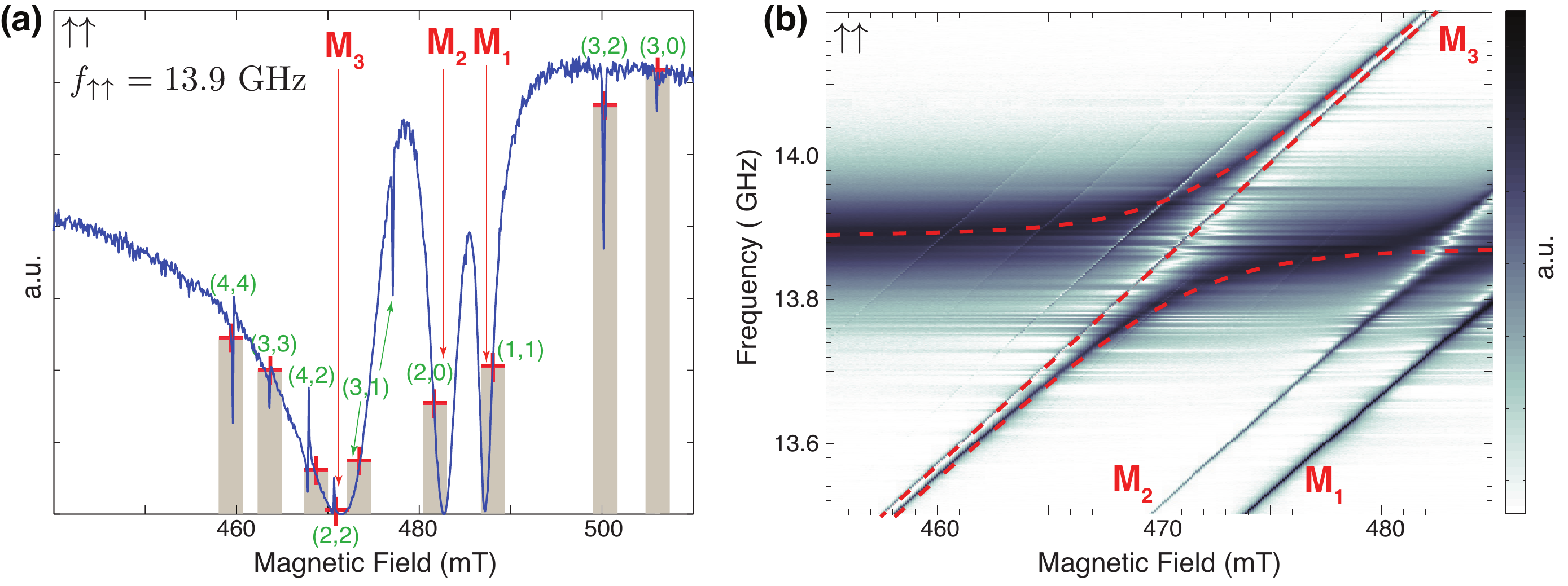}
	\caption{(a) Magnon modes observed near the dark cavity mode $\uparrow\uparrow$ at $f_{\uparrow\uparrow}=13.9$~GHz. Red crosses and shaded areas are theoretical predictions for magnon modes of order $(n,m)$\cite{fletcher}, (b) interaction between the magnon doublet M$_3$ and the dark cavity mode. Avoided crossings between photon and magnon modes are observed as transmission dips. The dashed curves show the three-mode interaction model fit.}
	\label{mmodes}
\end{figure*}

For the `dark' resonance, the strongest interaction is achieved for the magnon mode M$_3$ at at $B=$ 0.471~T and is demonstrated in the sub-figure labelled $\uparrow\uparrow$. The field of dynamic magnetization for the M$_1$ magnon mode is perfectly uniform over the volume of the sphere, but the microwave magnetic field of the dark cavity mode is perfectly antisymmetric with respect to the middle of the distance between the cavity posts. Thus, the M$_1$ mode does not interact with the ``dark" cavity mode due {to the resonant magnetic rf field reversing direction within the sphere, which causes first order cancellation of the coupling.}  The strong coupling to the M$_3$ mode suggests it has two variations of the magnon phase in the azimuthal direction\cite{dotsch}.
Thus, it is clear from these results that one must take into account the mode shapes of both the photonic and magnon resonances inside the sphere to properly calculate the coupling for a general photon-magnon mode interaction. Various magnon modes of the sphere can be observed at the lower frequency cavity resonance as depicted in Fig.~\ref{mmodes}(a). The figure also presents theoretical predictions made based on the existing theory of magnon modes in ferrimagnetic spheroids\cite{fletcher} with a fitted value of saturation magnetisation $M=0.255$~T for various combinations of wave numbers $(n,m)$. Here, for simplicity, we employ a simplified two-index notation dropping the third number referring to the number of the solution\cite{fletcher}. This estimation of the saturation magnetisation at $20$~mK is in good agreement with previous measurements\cite{Solt:1962lh} where it was demonstrated that $M$ grows from 0.175~T at 300 K to 0.244~T at 4 K. Visualisations of magnon modes in a sphere have been computed in the past \cite{dotsch}. The only parameter of the model, $M$, is fitted in such a way that the observed mode structure matches the predicted one. Other theoretically predicted modes have frequencies above the analysed frequency range and cannot be coupled to due to mode symmetries.
 
The strong coupling regime between the dark cavity mode and the M$_3$ magnon resonance is achieved at $B=0.471$~T, shown in Fig.~\ref{mmodes}(b). This plot displays a central peak representing the bare cavity resonance $f_{\uparrow\uparrow}$ outside the interaction, having linewidth $2\delta_{\uparrow\uparrow}$, as well as the splitting between the two peaks in the strong coupling regime (linewidths $2\delta_{\uparrow\downarrow+}$, $2\delta_{\uparrow\downarrow-}$ and $2\delta_{\uparrow\downarrow0}$). Unlike the case of the magnon mode M$_1$, which interacts according to a two-mode model with the bright mode, this interaction can only be successfully modelled using a three-mode model, one of which is the dark cavity mode, and the other two of which are a magnon mode doublet with the degeneracy lifted.
 
The origin of the double magnon mode in a spherical geometry has certain similarities with the existence of Whispering Gallery Mode (WGM) doublets that are extensively observed in spherical, toroidal and cylindrical resonators. The latter phenomenon is related to the existence of counter-propagating photonic modes with the same wave number whose degeneracy is lifted by the existence of backscattering elements. In the present experiment, a doublet of magnon resonances (magnetic in nature) exists where one is strongly coupled to a single photonic cavity resonance, and the second remains unaffected by the interaction. This is opposite to the case in which WGMs interact with paramagnetic impurities in high-$Q$ resonators, where the doublet is instead photonic in nature and interacts with a single (para)magnetic resonance. In this case, one resonance of the WGM doublet couples strongly to the electron spin resonance, and the other does not\cite{karim1, PhysRevA.89.013810,Goryachev:2014aa}.
 
The magnon linewidths can be estimated by direct measurements of the cavity transmission far from cavity-magnon interactions.
The linewidths of the uncoupled uniform magnon mode M$_1$ and mode M$_2$ are measured to be approximately $2\delta_{M1}=$ 1.1~MHz and $2\delta_{M2}=$ 760~kHz. The doublet magnon mode M$_3$ exhibits linewidths of 1.2~MHz and 490~kHz for the interacting and non-interacting resonances respectively. The sub-MHz line width of the YIG magnon resonances place it in front of the many other well-studied dilute spin systems such as Er$^{3+}$ and Eu$^{3+}$ and in Y$_2$SiO$_{5}$ ($\sim7-24$~MHz)\cite{Probst:2013zg}, Cr$^{2+}$ and Fe$^{3+}$ in sapphire ($9$ and $27$~MHz)\cite{PhysRevB.88.224426}, Er$^{3+}$ in YAlO$_3$ ($\sim15-33$~MHz)\cite{PhysRevB.90.075112}, etc, but behind such system as NV centre in diamond\cite{PhysRevLett.107.060502}. 
Traditionally the loss rate in resonant spin systems is expressed as an applied-field resolved resonance linewidth (half-width at half-maximum).
For the sphere in this work, the resonance linewidth is 0.2 Oersted as specified by the manufacturer and is confirmed by the present measurements. By multiplying this value by the gyromagnetic coefficient extracted from our experimental data one obtains the frequency resolved resonance linewidth slightly below 1 MHz, which is consistent with our results.
It should be mentioned that the single-crystal YIG is the material with lowest known ferromagnetic loss at microwave frequencies, far exceeding other ferri- or ferromagnetic materials. Furthermore, the shape of a (small) sphere ensures the highest possible symmetry for the collective dynamics of strongly exchange-coupled spins. This further decreases magnetic losses by prohibiting several processes leading to resonance linewidth broadening. These processes include a number of magnon-magnon scattering processes, as well as losses due to the excitation of higher-order resonances and travelling spin waves. Higher-order magnon modes observed in the experiment are absent at room temperature, but their appearance at cryogenic temperatures may be explained by mechanical stress induced in the sphere during the cool down procedure which slightly breaks the symmetry of the spin dynamics.


The cavity linewidths are 33 and 27~MHz  corresponding to $Q$-factors of 520 and 714 respectively for the dark and bright modes. Using these results and the simulated values of the geometric factor $G$, we estimate the effective surface resistance of the cavity to be 76~m$\Omega$. It should be noted that the cavity was fabricated from the OFHC copper and was not optimised in terms of loss. In particular, the inner surfaces of the cavity had not been polished and were oxidised. The surface resistance of ultra-pure polished copper at millikelvin temperatures is about 9~m$\Omega$, which could result in a considerable reduction of cavity bandwidths. Due to comparatively high cavity losses, the system demonstrates no nonlinear effects around the working incident power. The number of cavity photons is estimated to be less than 15 on resonance from the known incident power to the cavity, coupling and $Q$-factor\cite{Hartnett:2011aa,DanAPL}
 

\section*{Photon-Magnon Interaction}

The interaction between the bright cavity mode and the uniform magnon mode of the sphere can be described by the Hamiltonian of a harmonic oscillators (HO) coupled to an ensemble of spins\cite{PhysRevLett.104.077202,Soykal:2010ly,PhysRevB.82.104413}:
\begin{multline}
	\label{B004SFa}
	H_{\uparrow\downarrow M_1}=
	\hbar\omega_{\uparrow\downarrow} a^\dagger a+\frac{g\mu_B}{2}BS_z
	-\hbar g_{\uparrow\downarrow}(S_+a+a^\dagger S_-)\\
	+H_\text{nonJC}(aS^-,a^\dagger S^+),\\
\end{multline}
where $g$ is the electron $g$-factor, $\mu_B$ is the Bohr magneton, $B$ is the applied DC magnetic field, $a^\dagger$ and $a$ are the creation and annihilation operators for the bright cavity mode, $\omega_{\uparrow\downarrow}$ is the corresponding angular frequency, 
 $g_{\uparrow\downarrow}$ are coupling constants of the modes to the spins, and $S^z$ ($S^+$ and $S^-$) are the collective spin operators. The latter are operators whose eigenstates $\Ket{l,m}$ are similar to that in the Dicke model\cite{Dicke:1954fp}, i.e. $S_\pm\Ket{l,m}=\sqrt{(l\mp m)(l\pm m+1)}\ket{l,m\pm1}$. Note that the final term of the Hamiltonian is the typically oscillating ``counter-rotating" term, and is always neglected when using the rotating-wave approximation.
 
The first term of the Hamiltonian represents a HO corresponding to the bright cavity mode. The second term represent a collective spin with the frequency $\omega=\frac{{g\mu_B}}{\hbar}B$, and the last term determines the coupling between the cavity and the uniform magnon mode. 
From Eq.~\ref{B004SFa} it follows that one can use the  standard model of two interacting HOs to fit the experimental data. This fit gives the coupling constant $g_{\uparrow\downarrow}/\pi = $ 2.05~GHz or $2g_{\uparrow\downarrow}/\omega_{\uparrow\downarrow}\approx$ 10\%.
 
\begin{figure}[t!]
			\includegraphics[width=0.48\textwidth]{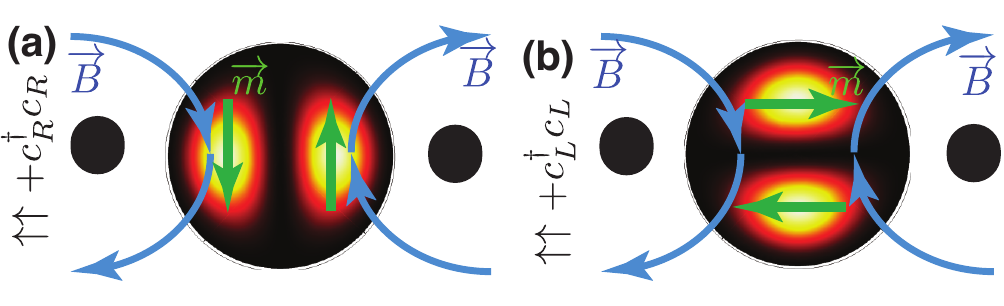}
	\caption{Origins of the different couplings of the dark cavity mode to two orthogonal realisations ($c_R^\dagger c_R$ (a) and $c_L^\dagger c_L$ (b) in Eq.~\ref{B005SFa}) of the M3 mode which is (2,2) magnon sphere mode. The plot shows the top cavity view with two black circles denoting the post orientation with respect to the sphere.}
	\label{couplings}
\end{figure}

As discussed previously, the interaction between the dark cavity mode and magnon mode M$_3$ is essentially a three-mode interaction. This follows from the fact that the observed interaction cannot be approximated by the two mode model, and is also confirmed by the existence of the degeneracy in the M3 mode. The corresponding fit of the three-mode model interaction is shown in Fig.~\ref{mmodes}(b).  This density plot demonstrates a central magnon line not interacting with an avoided crossing which it passes. As a result, one observes simultaneously three resonant peaks at a cross section which passes the centre of the avoided crossing. Since the upper and lower parts of the avoided crossing do not converge to the same asymptote, the system cannot be fit by the two oscillator model. The difference between the upper and  lower asymptotes defines the coupling between the two almost degenerate magnon modes. Such a three mode model has been previously utilzed for high-$Q$ Whispering Gallery Mode systems to fit interactions between two centre-propagating near-degenerate photon modes and a spin ensemble\cite{PhysRevA.89.013810,Goryachev:2014aa}. Thus, the three mode system appears as follows
\begin{multline}
	\label{B005SFa}
	H_{\uparrow\uparrow M_3}=
	\hbar\omega_{\uparrow\uparrow} b^\dagger b+\Big(\frac{g\mu_B}{2}B+\omega_R\Big)c_R^\dagger c_R\\
	\displaystyle+\Big(\frac{g\mu_B}{2}B+\omega_L\Big)c_L^\dagger c_L+\hbar g_{RL}(c_Rc_L^\dagger+c_R^\dagger c_L)
	\\-\hbar g_{\uparrow\uparrow}(c_R^\dagger a+a^\dagger c_R),
\end{multline}
 where $\omega_R$ and $\omega_L$ characterise the two doublet frequencies of the M$_3$ magnon mode, with creation (annihilation) operators $c_R^\dagger$ ($c_R$) and $c_L^\dagger$ ($c_L$) coupled via a backscattering mechanism with strength $g_{RL}$. As follows from this fit (dashed curves in Fig.~\ref{mmodes}(b)), only one of the counter-propagating modes of the doublet is coupled to the photon cavity resonance, with strength $g_{\uparrow\uparrow}$. {As shown in Fig.~\ref{mmodes}, M3 is the (2,2) sphere magnon mode, which implies the existence of a spatially orthogonal degenerate doublet. 
Figure~\ref{couplings} captures the origin of the difference in couplings to this doublet: whereas one realisation of the (2,2) mode maximises overlapping between the cavity RF magnetic field $\overrightarrow{B}$ and sphere RF magnetisation $\overrightarrow{m}$, the other minimises this quantity.}
The fit to the three mode HO model given in Eq.~\ref{B005SFa} gives two coupling constants $g_{\uparrow\uparrow}/\pi \approx$ 143~MHz and $g_{RL}/\pi\approx$ 12.5~MHz. Since $g_{\uparrow\uparrow}/\pi\gg \frac{1}{2}(\delta_{\uparrow\uparrow}+\delta_{M3})$,
 this interaction also satisfies the condition of the strong coupling regime $\frac{g_{\uparrow\uparrow}}{\pi}\gg \delta_{\uparrow\uparrow+}+\delta_{\uparrow\uparrow-}$ as demonstrated in the $\uparrow\uparrow$ plot of Fig.~\ref{density}. The corresponding spin cooperativities $\frac{g_{\uparrow x}^2}{(\pi)^2\delta_M\delta_{\uparrow x}}$ for the dark and bright modes are $1.6\times10^3$ and $1.3\times 10^5$ respectively. 
 
The predicted filling factors may be related to actual measurement of the magnon-photon coupling as 
$
	 {g_{\uparrow x}} = \omega_{\uparrow x}\sqrt{\chi\xi_{\uparrow x}}
$
where $\chi$ is the effective susceptibility of the YIG, $\omega_{\uparrow x}$ is either $\omega_{\uparrow\downarrow}$ or $\omega_{\uparrow\uparrow}$, and $\xi_{\uparrow x}$ is the corresponding calculated filling factor. The ratio between magnon-photon couplings is thus given as follows:
\begin{equation}
	\label{B006SFa}
	\displaystyle  \frac{g_{\uparrow \downarrow}}{g_{\uparrow \uparrow}}= \frac{\omega_{\uparrow \downarrow}}{\omega_{\uparrow \uparrow}}\sqrt{\frac{\xi_{\uparrow \downarrow}}{\xi_{\uparrow \uparrow}}}.\\
\end{equation}
The left-hand side of this expression estimated from the actual experiment is 14.5, whereas the right-hand found from the modelling is 15.0, demonstrating good agreement between modelling and experiment. Further optimisation of the cavity dimensions, as has already been discussed, should result in an increase of the coupling strength to $5.2$~GHz, representing about 24\% of the total cavity energy. Such an optimised system is characterised by a spin cooperativity of $9.1\times 10^5$, an increase over the current cooperativity by a factor of 7 assuming that the cavity parameters remain the same.

Assuming spin density in YIG to be \hbox{$2.1\times 10^{22}$~cm$^{-3}$}\cite{PhysRevLett.113.083603}, the coupling per spin achieved in this work can be estimated to be $0.3$~Hz. This value is almost an order of magnitude larger than previously achieved $0.038$~Hz in a microwave cavity\cite{PhysRevLett.113.083603} and considerably larger than for a millimetre-wave cavity $0.067$~Hz\cite{yig2}. As for the latter work, similar results in terms of coupling strength have been achieved by scaling down the size of the cavity leading to an increase of resonant frequencies. However, a shift to the millimetre-wave frequency range leads to additional complications related to higher loss and requirement for larger magnetic fields.

\section*{Predictions for the Optimised Cavity}

\begin{figure}[t!]
	\centering
			\includegraphics[width=0.48\textwidth]{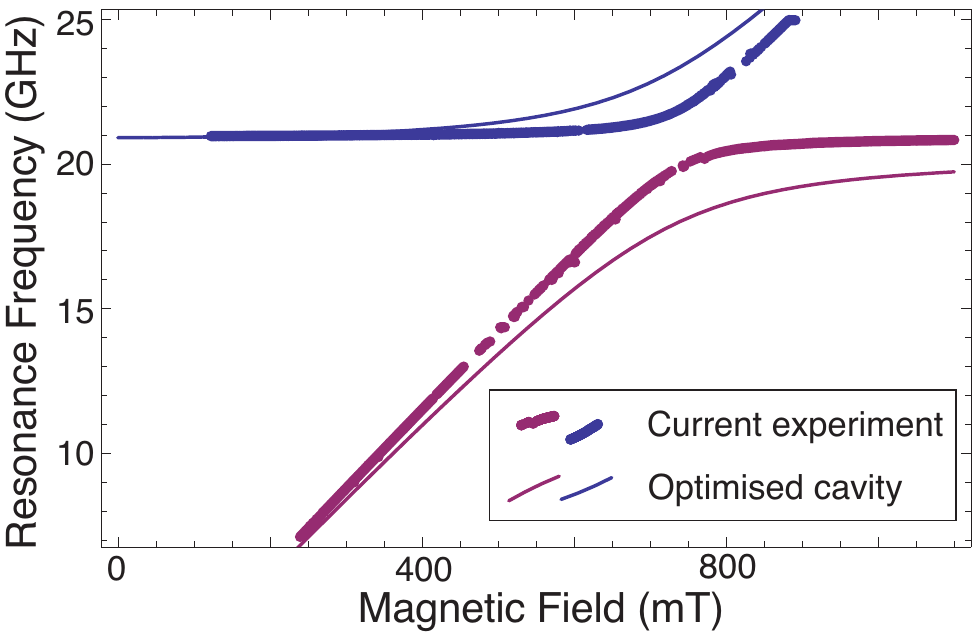}
	\caption{Predictions of the optimised system response with the cavity tuned to the resonant frequency observed in the experiment.}
	\label{predic}
\end{figure}

The results obtained in this work are preliminary and there is a large potential for further improvement. In particular, in order to decrease the cavity frequency such that magnon-photon interactions can occur in the low gigahertz range and at low DC magnetic field, the post gap size can simply be reduced.  Gap sizes on the order of micrometers to tens of micrometers are easily obtained, which will allow the tuning of such cavities to a few gigahertz. With more advanced manufacturing technology, we anticipate that reduction of the gap size could be even further reduced to the nanoscale. An alternative approach to reduce the cavity frequency is to utilize a more complicated post structure\cite{mpost,patent2014} or to fill the cavity gap with high-permittivity dielectric which increases the equivalent capacitance. Here, we only predict the system response when the cavity parameters are optimised using realistic values, easily obtainable in the next design iteration. As discussed in the paper, Fig.~\ref{predic} shows the system response with a predicted photon-magnon coupling of $5.2$~GHz for the bright photonic mode. Here, the asymmetry of the interaction is enhanced which is a property of the ultra-strong coupling regime. The dark mode exhibits similar enhancement, but the coupling strength always stays an order of magnitude lower.  

In addition to the enhancement of the coupling, further optimisation will lead to an increase in the geometric factor by up to a factor of two. This, together with applying standard surface polishing and surface treatment techniques to improve the surface resistance of the cavity, should see at least a 12 fold reduction of the cavity linewidth. These improvements potentially lead to a very large enhancement of the cooperitivity of the photon-magnon coupling in this type of a microwave cavity. From these estimations, a value of $1.1 \times 10^7$ should be possible, which is almost a factor of 100 times better than the results reported in this work. This would correspond to a coupling-per-spin as high as $0.77$~Hz.

\hspace{10pt}

In conclusion, we have demonstrated strong coupling of 2~GHz and high cooperativity of $10^5$ between a photonic mode of a field-focusing double post microwave reentrant cavity and magnon resonances of a sub-millimetre size YIG sphere. The novel cavity design allowed the magnetic filling factor within the small spherical sample to be much larger than possible using standard cavity techniques. The same cavity allowed much lower cavity resonant frequencies to be achieved than the resonant frequencies of the sample itself. The unique properties of the cavity allows us to better utilise high spin density of the YIG crystal for cavity QED experiments. 

\hspace{15pt}

\section*{Acknowledgements}

This work was supported by the Australian Research Council Grant No. CE110001013, FL0992016 and DP110103980. 

\section*{References}
%

\end{document}